\documentclass[aip,apl]{revtex4-1}

\usepackage{graphicx}
\usepackage{dcolumn}
\usepackage{bm}
\usepackage{amsmath}
\usepackage{latexsym}

\newcommand{\rfig}[1]{Fig.~\ref{#1}}
\newcommand{\rref}[1]{Ref.~\cite{#1}}

\begin{document}

\title{Half integer quantum Hall effect in high mobility single layer epitaxial graphene}

\author{Xiaosong Wu$^1$, Yike Hu$^1$, Ming Ruan$^1$, Nerasoa K Madiomanana$^1$, John Hankinson$^1$, Mike Sprinkle$^1$, Claire Berger$^{1,2}$, Walt A. de Heer$^{1}$}
\affiliation{
$^1$School of Physics, Georgia Institute of Technology, Atlanta, GA 30332 \\
$^2$CNRS- Institut N\'eel,  BP 166, 38042 Grenoble Cedex 9, France}


\begin{abstract}
The quantum Hall effect, with a Berry's phase of $\pi$ is demonstrated here on a single graphene layer grown on the C-face of 4H silicon carbide. The mobility is $\sim$ 20,000 cm$^2$/V$\cdot$s at 4 K and ~15,000 cm$^2$/V$\cdot$s at 300 K despite contamination and substrate steps. This is comparable to the best exfoliated graphene flakes on SiO$_2$ and an order of magnitude larger than Si-face epitaxial graphene monolayers. These and other properties indicate that C-face epitaxial graphene is a viable platform for graphene-based electronics.
\end{abstract}

\maketitle

In 2004 Berger {\it et al.} introduced the concept of graphene-based electronics and a route to realize it \cite{Berger2004}. Results in that paper on patterned epitaxial graphene grown on the Si-face of 4H silicon carbide crystals showed that the material could be top-gated and it also showed magneto-resistance measurements with characteristic Shubnikov de Haas oscillations (SdHOs). In one sample (sample A of \rref{Berger2004}), the SdHOs corresponding to the $n=2$ and $n=3$ Landau levels were observed as well as a weak modulation in the Hall effect.  The quantum Hall effect (QHE) was not observed, at least not for these relatively large index Landau levels. The mobility was also relatively low ($\mu$=1100 cm$^2$/V$\cdot$s).  From LEED and AES we estimated that the material was 3 monolayers, but we now know, from better understanding of the structure  \cite{Hass2008} and the low field linearity of the Hall effect, that it was in fact a single layer. We also have found that relatively low mobilities are characteristic for Si-face epitaxial graphene compared with the C-face.

Subsequent work focused on multilayered epitaxial graphene (MEG) grown on the carbon face of silicon carbide which demonstrated high moblities.To compare with previous work a summary of the properties of MEG is provided, although the experiments were performed on single layer epitaxial graphene (SEG).

In epitaxial graphene, the graphene layer at the interface with the substrate is  n-doped due to the Schottky barrier at the interface, while the other layers are essentially undoped. The individual layers of MEG are electronically decoupled due to an unusual rotational order  \cite{Hass2008} and consequently  the layers exhibit graphene properties rather than graphitic properties as might be expected \cite{Berger2006,Sadowski2006,Wu2007,Orlita2008,Plochocka2008,Sun2008, Miller2009,Sprinkle2009}. Specifically, the interface layer, which carries most of the current, shows the non-trivial Berry's phase and has a high mobility ($>$30,000 cm$^2$/V$\cdot$s) while the other, undoped layers have even much higher mobilities ($>$250,000 cm$^2$/V$\cdot$s). Moreover, at least the top-most layer has been shown to be continuous over the entire surface, without breaks or crystallographic domain boundaries\cite{Miller2009}. Nevertheless, the quantum Hall effect (QHE) is not seen \cite{ Darancet2008,deHeer2007}. 

Many applications, including THz electronics \cite{Moon2009}, require room temperature mobilities of the order of 10,000 cm$^2$/V$\cdot$s. We have demonstrated patterned samples with $\mu\sim$30,000 cm$^2$/V$\cdot$s in multilayer epitaxial graphene grown on the C-face of hexagonal SiC (MEG) using the so-called RF furnace method \cite{Berger2006,deHeer2007,Hass2008b}, in which the substrate is enclosed in a graphitic chamber with or without inert gas to produce high quality graphene  of arbitrary thickness. Although challenging we succeeded in producing SEG on the C-face of SiC (it is easier to produce SEG on the Si-face \cite{Emtsev2009,Virojanadara2008}). 

\begin{figure}
\includegraphics[width=0.7\textwidth]{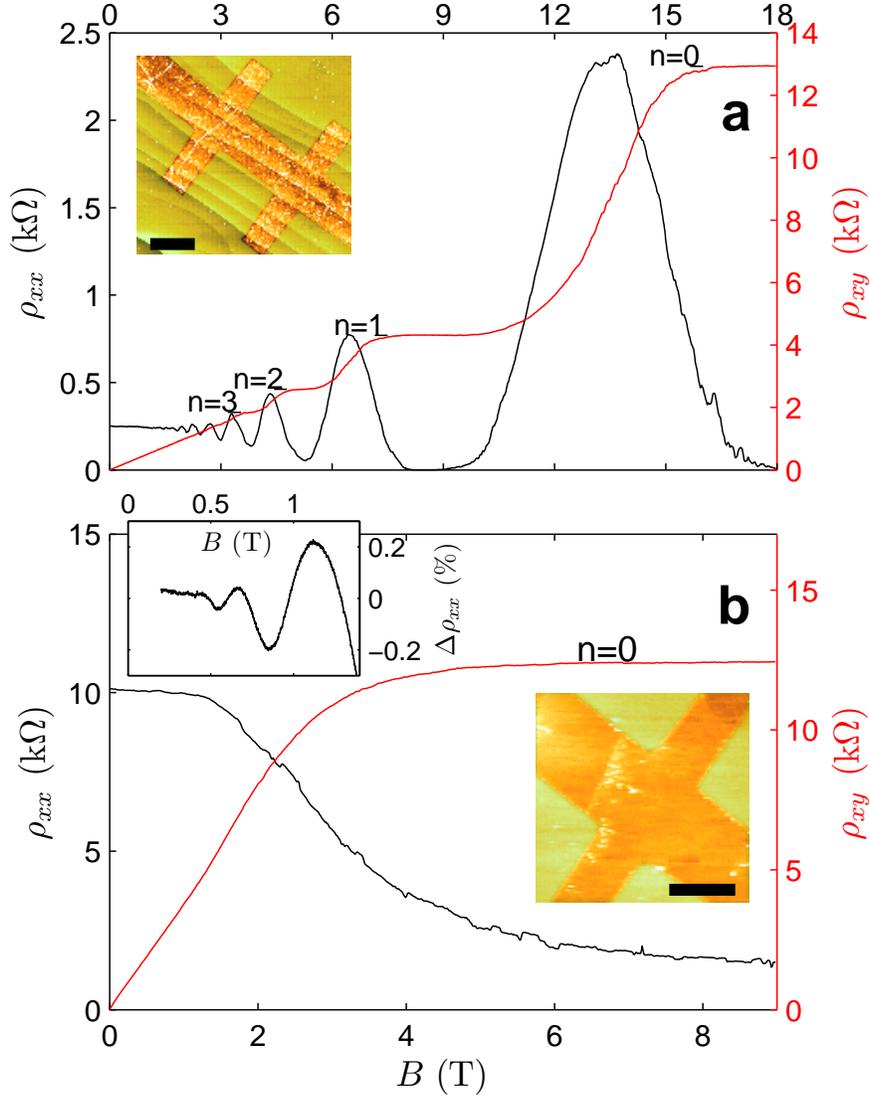}
\caption{\label{fig:QHE} Quantum Hall effect in two monolayer epitaxial graphene samples measured at 1.4 K. Panel {\bf a}, Sample A. Hall resistance (red) as a function of magnetic field showing characteristic Hall plateaus at $\rho_{xy}=(h/4e^2)/(n+1/2)$ where $n$ is the Landau level index, magnetoresistivity $\rho_{xx}$ (black) showing characteristic oscillations and droping to zero for low Landau indexes. Intriguing reproducible fine structure features are observed in both $\rho_{xx}$ and $\rho_{xy}$ in high fields. Inset: AFM image of the Hall bar (1.8 $\mu$m $\times$ 4.6 $\mu$m) patterned over several SiC steps, showing e-beam resist residue particles (white spots, covering about 17$\%$ of the surface) and pleats in the graphene (white lines).  Panel {\bf b}, Sample B. Red: Hall resistance showing a Hall plateau for $n=0$. Black: magnetoresistance $\rho_{xx}$ . Very weak oscillations can be discerned at $n=1,2,3$ (upper inset). Lower Inset: AFM image of the Hall cross (1.5 $\mu$m $\times$ 2.5 $\mu$m). The substrate is step-free and the surface is clean. The scale bars represent 2 $\mu$m.}
\end{figure}

Here we show the QHE in SEG in two samples (sample A and sample B). The inset of \rfig{fig:QHE}{\bf a} shows an atomic force microscopy image (AFM) of sample A, a Hall bar that was deliberately patterned \rref{Berger2006} over steps on the substrate surface in order to evaluate the effects of steps on the transport properties. The AFM image shows e-beam resist residue particles from the processing and characteristic pleats (white lines) on the graphene surface  \cite{Hass2008}.
 
The mobility of the sample is 20,000 cm$^2$/V$\cdot$s. The QHE is well resolved in \rfig{fig:QHE}{\bf a}, which shows quantum Hall plateaus in the magnetic field dependence of the Hall resistance, as first observed in exfoliated graphene flakes on SiO$_2$ (EGF) \cite{Zhang2005,Novoselov2005}. The Hall plateaus correspond to transverse resistances $\rho_{xy}=(h/4e^2)/(n+1/2)$ for $n$=0 to 3, where $n$ is the Landau level index, which establishes the non-trivial Berry's phase of $\pi$\cite{Ando1998a}. The longitudinal resistivity $\rho_{xx}$ (\rfig{fig:QHE}{\bf a}) shows the characteristic SdHOs, in which Landau levels from $n=0$ up to $n=8$ are easily recognized.  The SdHOs develop into the QHE in high fields, manifested by characteristic zero resistance minima and Hall plateaus.  The graphene charge density obtained from the Hall coefficient (temperature independent) is found to be $n_s=1.27\times10^{12}$ /cm$^2$ (hole doped) and temperature independent. The graphene layer is negatively doped due to the work function difference at the SiC graphene interface \cite{Berger2006}. 

The graphene surface has accumulated its positive charge from environmental humidity. The charge density can be controlled by adjusting the exposure to humidity as well as by exposure to ambient light.  Note that epitaxial graphene surfaces can be immaculately cleaned by heating in vacuum to 1000 $^\circ$C \cite{Miller2009}. Also the conventional (local) top gating methods used for applications \cite{Kedzierski2008} cannot be used to demonstrate the QHE.

The second Hall bar (sample B) was deliberately patterned on a step-free terrace, \rfig{fig:QHE}{\bf b}; its mobility is 4000 cm$^2$/V$\cdot$s. It also exhibits the QHE, however only a single quantum Hall plateau corresponding to the n=0 Landau level is observed. The Shubnikov de Haas oscillations corresponding to the other Landau levels are very small and reminiscent of those seen in MEG \cite{deHeer2007,Wu2007}.

Sample A was measured three times after re-exposing it to conditions with different humidity. The charge densities $n_s$ are 0.9, 1.28 and 1.27 /cm$^2$, while the mobility varies slightly ($\sim 5\%$ cm$^2$/V$\cdot$s at 4.2 K.). The QHE is observed for all three experimental runs. 

Despite the fact that the graphene in sample A is draped over several steps, is heavily contaminated ({\it cf.} the hole doping and the particles) and has pleats, the mobility is as high as 20,000 cm$^2$/Vs at 4.2 K and 15,000 cm$^2$/ V$\cdot$s at 300 K and shows only a mild temperature dependence (similar to MEG samples \cite{Berger2006}). These observations show that 1. scattering from impurities is weak, 2. electron-phonon scattering is suppressed \cite{Orlita2008}, 3. the graphene is continuous over steps in the SiC substrate. 

\begin{figure}
\centering
\includegraphics[width=0.8\textwidth]{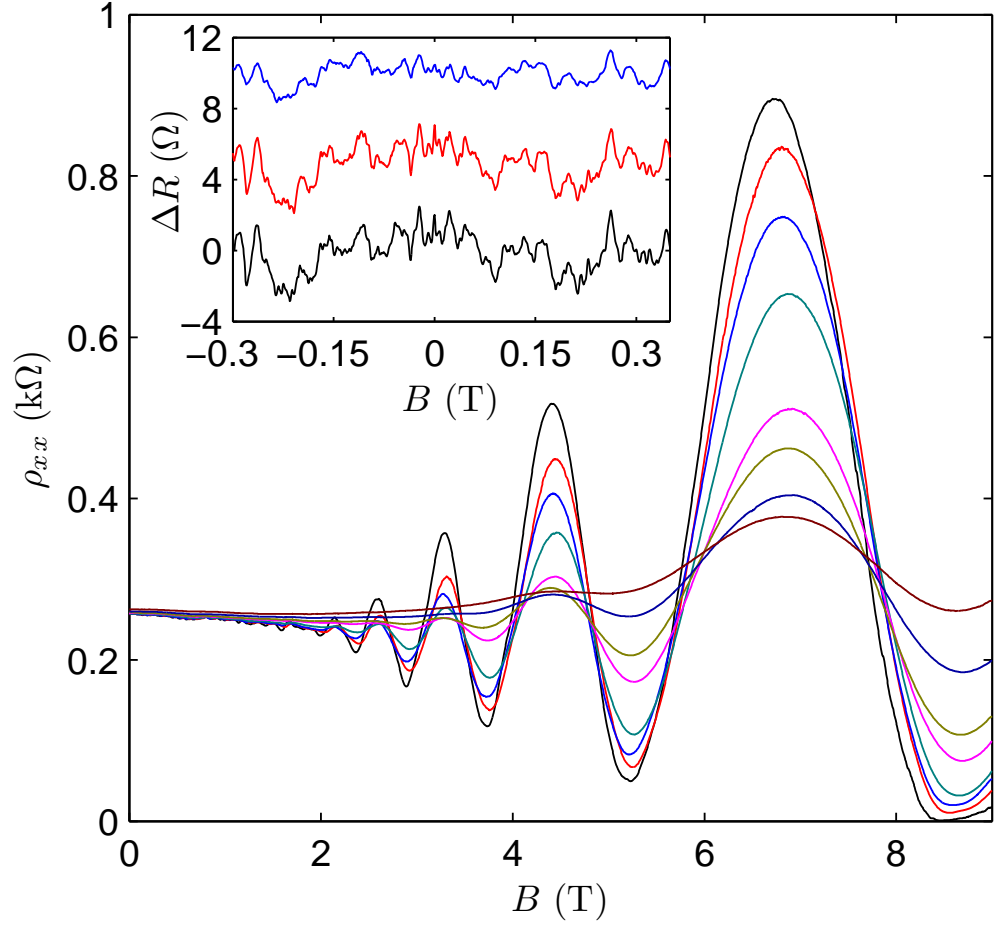}
\caption{\label{fig:SdHOsUCFs} Temperature dependence of the magnetoresistance. Panel {\bf a}, SdHOs at different temperatures, 1.4 K, 10 K, 20 K, 30 K, 50 K, 60 K, 80 K and 100 K. Inset: low field magnetoresistance at 4.2 K, 10 K and 30 K exhibits universal conductance fluctuations, indicating that the phase coherence length $l_\phi$ is comparable to the size of the sample. Data are shifted in $y$ axis for clarity.}
\end{figure}

For the case where the charge density $n_s$ is 1.28$\times 10^{12}$ /cm$^2$, the experiment has been carried out in magnetic fields up to 9 T and at temperatures up to 150 K, where SdHOs for the $n=1$ Landau level can still be seen. The temperature dependence of SdHOs are plotted in \rfig{fig:SdHOsUCFs}. The damping of the oscillations with temperature is caused by thermal broadening of Landau levels. In graphene, the temperature dependence of the amplitude is described by $t_k/\sinh t_k$, where $t_k=2\pi^2k_BT\mu/\hbar v_0^2eB$ \cite{Gusynin2005a}. Here, $\mu$, $T$, $B$ and $v_0$ are chemical potential, temperature, magnetic field and the band velocity, respectively. Using this formula, we find the velocity $v_0=1.14\times10^6$ m/s, which agrees with the $v_0$ of graphene flakes on SiO$_2$  \cite{Zhang2005,Novoselov2005}, which, combined with the graphene Berry's phase establishes that the SiC substrate does not affect the properties of EG any more than the SiO$_2$ substate affects exfoliated graphene.

We have also investigated the low field magnetoresistance. As shown in the inset of \rfig{fig:SdHOsUCFs}, the sample displays aperiodic and reproducible universal conductance fluctuations (UCFs), that diminish with increasing temperatures. The phase coherence length  is estimated from the magnetoconductance correlation function: $F(\Delta B)= \langle(G(B)-\langle G(B)\rangle)(G(B+\Delta B)-\langle G(B+\Delta B)\rangle)\rangle$ \cite{Beenakker1991}. A correlation field $\Delta B_c$ is defined as the half-width at half-height $F(\Delta B_c)=F(0)/2$. The phase coherence length $l_\phi$ is related to $\Delta B_c$ by $\Delta B_c\approx e/hl_\phi^2$. We find $l_\phi \approx$ 0.6 $\mu$m at 4.2 K, similar to values previously found in MEG ribbons \cite{Berger2006,Wu2007}.

Besides the QHE, we have shown that 1.  single layer epitaxial graphene can be grown on the C-face of hexagonal silicon carbide wafers; 2.  the graphene sheet is continuous over substrate steps; 3. its mobility rivals that of the best exfoliated graphene on SiO$_2$, despite significant contamination, substrate steps, and harsh processing procedures; 4. the QHE in C-face epitaxial graphene demonstrates that the substrate is at least as unimportant here as it is for exfoliated graphene on SiO$_2$. 

Concluding, the robustness and large scale patterning that is possible with epitaxial graphene opens new avenues for graphene physics.  This important development brings epitaxial graphene yet a step closer to becoming a scalable platform for graphene-based electronics as anticipated \cite{Berger2004,deHeer2006}.

This work was supported by NSF grant DMR-0820382 and the W. M. Keck Foundation. A portion of this work was performed at the National High Magnetic Field Laboratory, which is supported by NSF Cooperative Agreement No. DMR-0654118, by the State of Florida, and by the DOE. We would like to thank Z.~G. Jiang for insightful discussions. We also acknowledge E.~C. Palm, T.~P. Murphy, J.-H. Park, G.~E. Jones and Z.~L. Guo for experimental assistance.

\begin{thebibliography}{23}
\expandafter\ifx\csname natexlab\endcsname\relax\def\natexlab#1{#1}\fi
\expandafter\ifx\csname bibnamefont\endcsname\relax
  \def\bibnamefont#1{#1}\fi
\expandafter\ifx\csname bibfnamefont\endcsname\relax
  \def\bibfnamefont#1{#1}\fi
\expandafter\ifx\csname citenamefont\endcsname\relax
  \def\citenamefont#1{#1}\fi
\expandafter\ifx\csname url\endcsname\relax
  \def\url#1{\texttt{#1}}\fi
\expandafter\ifx\csname urlprefix\endcsname\relax\def\urlprefix{URL }\fi
\providecommand{\bibinfo}[2]{#2}
\providecommand{\eprint}[2][]{\url{#2}}

\bibitem[{\citenamefont{Berger et~al.}(2004)\citenamefont{Berger, Song, Li, Li,
  Ogbazghi, Feng, Dai, Marchenkov, Conrad, First et~al.}}]{Berger2004}
\bibinfo{author}{\bibfnamefont{C.}~\bibnamefont{Berger}},
  \bibinfo{author}{\bibfnamefont{Z.~M.} \bibnamefont{Song}},
  \bibinfo{author}{\bibfnamefont{T.~B.} \bibnamefont{Li}},
  \bibnamefont{et~al.}, \bibinfo{journal}{J. Phys. Chem. B}
  \textbf{\bibinfo{volume}{108}}, \bibinfo{pages}{19912}
  (\bibinfo{year}{2004}).

\bibitem[{\citenamefont{Hass et~al.}(2008{\natexlab{a}})\citenamefont{Hass,
  Varchon, Mill\'{a}n-Otoya, Sprinkle, Sharma, de~Heer, Berger, First, Magaud,
  and Conrad}}]{Hass2008}
\bibinfo{author}{\bibfnamefont{J.}~\bibnamefont{Hass}},
  \bibinfo{author}{\bibfnamefont{F.}~\bibnamefont{Varchon}},
  \bibinfo{author}{\bibfnamefont{J.~E.} \bibnamefont{Mill\'{a}n-Otoya}},
  \bibnamefont{et~al.}, \bibinfo{journal}{Phys. Rev. Lett.}
  \textbf{\bibinfo{volume}{100}}, \bibinfo{pages}{125504}
  (\bibinfo{year}{2008}{\natexlab{a}}).

\bibitem[{\citenamefont{Berger et~al.}(2006)\citenamefont{Berger, Song, Li, Wu,
  Brown, Naud, Mayo, Li, Hass, Marchenkov et~al.}}]{Berger2006}
\bibinfo{author}{\bibfnamefont{C.}~\bibnamefont{Berger}},
  \bibinfo{author}{\bibfnamefont{Z.~M.} \bibnamefont{Song}},
  \bibinfo{author}{\bibfnamefont{X.~B.} \bibnamefont{Li}},
  \bibnamefont{et~al.}, \bibinfo{journal}{Science}
  \textbf{\bibinfo{volume}{312}}, \bibinfo{pages}{1191} (\bibinfo{year}{2006}).

\bibitem[{\citenamefont{Sadowski et~al.}(2006)\citenamefont{Sadowski, Martinez,
  Potemski, Berger, and de~Heer}}]{Sadowski2006}
\bibinfo{author}{\bibfnamefont{M.~L.} \bibnamefont{Sadowski}},
  \bibinfo{author}{\bibfnamefont{G.}~\bibnamefont{Martinez}},
  \bibinfo{author}{\bibfnamefont{M.}~\bibnamefont{Potemski}},
  \bibnamefont{et~al.}, \bibinfo{journal}{Phys. Rev. Lett.}
  \textbf{\bibinfo{volume}{97}}, \bibinfo{pages}{266405}
  (\bibinfo{year}{2006}).

\bibitem[{\citenamefont{Wu et~al.}(2007)\citenamefont{Wu, Li, Song, Berger, and
  de~Heer}}]{Wu2007}
\bibinfo{author}{\bibfnamefont{X.~S.} \bibnamefont{Wu}},
  \bibinfo{author}{\bibfnamefont{X.~B.} \bibnamefont{Li}},
  \bibinfo{author}{\bibfnamefont{Z.~M.} \bibnamefont{Song}},
  \bibnamefont{et~al.}, \bibinfo{journal}{Phys. Rev. Lett.}
  \textbf{\bibinfo{volume}{98}}, \bibinfo{pages}{136801}
  (\bibinfo{year}{2007}).

\bibitem[{\citenamefont{Orlita et~al.}(2008)\citenamefont{Orlita, Faugeras,
  Plochocka, Neugebauer, Martinez, Maude, Barra, Sprinkle, Berger, de~Heer
  et~al.}}]{Orlita2008}
\bibinfo{author}{\bibfnamefont{M.}~\bibnamefont{Orlita}},
  \bibinfo{author}{\bibfnamefont{C.}~\bibnamefont{Faugeras}},
  \bibinfo{author}{\bibfnamefont{P.}~\bibnamefont{Plochocka}},
  \bibnamefont{et~al.}, \bibinfo{journal}{Phys. Rev. Lett.}
  \textbf{\bibinfo{volume}{101}}, \bibinfo{pages}{267601}
  (\bibinfo{year}{2008}).

\bibitem[{\citenamefont{Plochocka et~al.}(2008)\citenamefont{Plochocka,
  Faugeras, Orlita, Sadowski, Martinez, Potemski, Goerbig, Fuchs, Berger, and
  de~Heer}}]{Plochocka2008}
\bibinfo{author}{\bibfnamefont{P.}~\bibnamefont{Plochocka}},
  \bibinfo{author}{\bibfnamefont{C.}~\bibnamefont{Faugeras}},
  \bibinfo{author}{\bibfnamefont{M.}~\bibnamefont{Orlita}},
  \bibnamefont{et~al.}, \bibinfo{journal}{Phys. Rev. Lett}
  \textbf{\bibinfo{volume}{100}}, \bibinfo{pages}{087401}
  (\bibinfo{year}{2008}).

\bibitem[{\citenamefont{Sun et~al.}(2008)\citenamefont{Sun, Wu, Divin, Li,
  Berger, de~Heer, First, and Norris}}]{Sun2008}
\bibinfo{author}{\bibfnamefont{D.}~\bibnamefont{Sun}},
  \bibinfo{author}{\bibfnamefont{Z.~K.} \bibnamefont{Wu}},
  \bibinfo{author}{\bibfnamefont{C.}~\bibnamefont{Divin}},
  \bibnamefont{et~al.}, \bibinfo{journal}{Phys. Rev. Lett.}
  \textbf{\bibinfo{volume}{101}}, \bibinfo{pages}{157402}
  (\bibinfo{year}{2008}).

\bibitem[{\citenamefont{Miller et~al.}(2009)\citenamefont{Miller, Kubista,
  Rutter, Ruan, de~Heer, First, and Stroscio}}]{Miller2009}
\bibinfo{author}{\bibfnamefont{D.~L.} \bibnamefont{Miller}},
  \bibinfo{author}{\bibfnamefont{K.~D.} \bibnamefont{Kubista}},
  \bibinfo{author}{\bibfnamefont{G.~M.} \bibnamefont{Rutter}},
  \bibnamefont{et~al.}, \bibinfo{journal}{Science}
  \textbf{\bibinfo{volume}{324}}, \bibinfo{pages}{924} (\bibinfo{year}{2009}).

\bibitem[{\citenamefont{Sprinkle et~al.}(2009)\citenamefont{Sprinkle, Siegel,
  Hu, Hicks, Soukiassian, Tejeda, Taleb-Ibrahimi, Fevre, Bertran, Berger
  et~al.}}]{Sprinkle2009}
\bibinfo{author}{\bibfnamefont{M.}~\bibnamefont{Sprinkle}},
  \bibinfo{author}{\bibfnamefont{D.}~\bibnamefont{Siegel}},
  \bibinfo{author}{\bibfnamefont{Y.}~\bibnamefont{Hu}}, \bibnamefont{et~al.}
  \bibinfo{journal}{Phys. Rev. Lett.} \textbf{\bibinfo{volume}{103}}, 
  \bibinfo{pages}{226803} (\bibinfo{year}{2009}).
  
  \bibitem[{\citenamefont{Darancet et~al.}(2008)\citenamefont{Darancet, Wipf,
  Berger, de~Heer, and Mayou}}]{Darancet2008}
\bibinfo{author}{\bibfnamefont{P.}~\bibnamefont{Darancet}},
  \bibinfo{author}{\bibfnamefont{N.}~\bibnamefont{Wipf}},
  \bibinfo{author}{\bibfnamefont{C.}~\bibnamefont{Berger}},
  \bibnamefont{et~al.}, \bibinfo{journal}{Phys. Rev. Lett.}
  \textbf{\bibinfo{volume}{101}}, \bibinfo{pages}{116806}
  (\bibinfo{year}{2008}).

\bibitem[{\citenamefont{Moon et~al.}(2009)\citenamefont{Moon, Curtis, Hu, Wong,
  McGuire, Campbell, Jernigan, Tedesco, VanMil, Myers-Ward et~al.}}]{Moon2009}
\bibinfo{author}{\bibfnamefont{J.}~\bibnamefont{Moon}},
  \bibinfo{author}{\bibfnamefont{D.}~\bibnamefont{Curtis}},
  \bibinfo{author}{\bibfnamefont{M.}~\bibnamefont{Hu}}, \bibnamefont{et~al.},
  \bibinfo{journal}{IEEE Electr. Device. L.} \textbf{\bibinfo{volume}{30}},
  \bibinfo{pages}{650} (\bibinfo{year}{2009}).

\bibitem[{\citenamefont{de~Heer et~al.}(2007)\citenamefont{de~Heer, Berger, Wu,
  First, Conrad, Li, Li, Sprinkle, Hass, Sadowski et~al.}}]{deHeer2007}
\bibinfo{author}{\bibfnamefont{W.~A.} \bibnamefont{de~Heer}},
  \bibinfo{author}{\bibfnamefont{C.}~\bibnamefont{Berger}},
  \bibinfo{author}{\bibfnamefont{X.~S.} \bibnamefont{Wu}},
  \bibnamefont{et~al.}, \bibinfo{journal}{Solid State Commun.}
  \textbf{\bibinfo{volume}{143}}, \bibinfo{pages}{92} (\bibinfo{year}{2007}).

\bibitem[{\citenamefont{Hass et~al.}(2008{\natexlab{b}})\citenamefont{Hass,
  de~Heer, and Conrad}}]{Hass2008b}
\bibinfo{author}{\bibfnamefont{J.}~\bibnamefont{Hass}},
  \bibinfo{author}{\bibfnamefont{W.~A.} \bibnamefont{de~Heer}},
  \bibnamefont{and} \bibinfo{author}{\bibfnamefont{E.~H.}
  \bibnamefont{Conrad}}, \bibinfo{journal}{J. Phys.: Condens. Matter}
  \textbf{\bibinfo{volume}{20}}, \bibinfo{pages}{323202}
  (\bibinfo{year}{2008}{\natexlab{b}}).

\bibitem[{\citenamefont{Emtsev et~al.}(2009)\citenamefont{Emtsev, Bostwick,
  Horn, Jobst, Kellogg, Ley, McChesney, Ohta, Reshanov, Rohrl
  et~al.}}]{Emtsev2009}
\bibinfo{author}{\bibfnamefont{K.~V.} \bibnamefont{Emtsev}},
  \bibinfo{author}{\bibfnamefont{A.}~\bibnamefont{Bostwick}},
  \bibinfo{author}{\bibfnamefont{K.}~\bibnamefont{Horn}}, \bibnamefont{et~al.},
  \bibinfo{journal}{Nat. Mater.} \textbf{\bibinfo{volume}{8}},
  \bibinfo{pages}{203} (\bibinfo{year}{2009}).

\bibitem[{\citenamefont{Virojanadara et~al.}(2008)\citenamefont{Virojanadara,
  Syv\"{a}jarvi, Yakimova, Johansson, Zakharov, and
  Balasubramanian}}]{Virojanadara2008}
\bibinfo{author}{\bibfnamefont{C.}~\bibnamefont{Virojanadara}},
  \bibinfo{author}{\bibfnamefont{M.}~\bibnamefont{Syv\"{a}jarvi}},
  \bibinfo{author}{\bibfnamefont{R.}~\bibnamefont{Yakimova}},
  \bibnamefont{et~al.}, \bibinfo{journal}{Phys. Rev. B}
  \textbf{\bibinfo{volume}{78}}, \bibinfo{pages}{245403}
  (\bibinfo{year}{2008}).

\bibitem[{\citenamefont{Zhang et~al.}(2005)\citenamefont{Zhang, Tan, Stormer,
  and Kim}}]{Zhang2005}
\bibinfo{author}{\bibfnamefont{Y.~B.} \bibnamefont{Zhang}},
  \bibinfo{author}{\bibfnamefont{Y.~W.} \bibnamefont{Tan}},
  \bibinfo{author}{\bibfnamefont{H.~L.} \bibnamefont{Stormer}},
  \bibnamefont{et~al.}, \bibinfo{journal}{Nature}
  \textbf{\bibinfo{volume}{438}}, \bibinfo{pages}{201} (\bibinfo{year}{2005}).

\bibitem[{\citenamefont{Novoselov et~al.}(2005)\citenamefont{Novoselov, Geim,
  Morozov, Jiang, Katsnelson, Grigorieva, Dubonos, and Firsov}}]{Novoselov2005}
\bibinfo{author}{\bibfnamefont{K.~S.} \bibnamefont{Novoselov}},
  \bibinfo{author}{\bibfnamefont{A.~K.} \bibnamefont{Geim}},
  \bibinfo{author}{\bibfnamefont{S.~V.} \bibnamefont{Morozov}},
  \bibnamefont{et~al.}, \bibinfo{journal}{Nature}
  \textbf{\bibinfo{volume}{438}}, \bibinfo{pages}{197} (\bibinfo{year}{2005}).

\bibitem[{\citenamefont{Ando et~al.}(1998)\citenamefont{Ando, Nakanishi, and
  Saito}}]{Ando1998a}
\bibinfo{author}{\bibfnamefont{T.}~\bibnamefont{Ando}},
  \bibinfo{author}{\bibfnamefont{T.}~\bibnamefont{Nakanishi}},
  \bibnamefont{and} \bibinfo{author}{\bibfnamefont{R.}~\bibnamefont{Saito}},
  \bibinfo{journal}{J. Phys. Soc. Jpn.} \textbf{\bibinfo{volume}{67}},
  \bibinfo{pages}{2857} (\bibinfo{year}{1998}).

\bibitem[{\citenamefont{Kedzierski et~al.}(2008)\citenamefont{Kedzierski, Hsu,
  Healey, Wyatt, Keast, Sprinkle, Berger, and de~Heer}}]{Kedzierski2008}
\bibinfo{author}{\bibfnamefont{J.}~\bibnamefont{Kedzierski}},
  \bibinfo{author}{\bibfnamefont{P.~L.} \bibnamefont{Hsu}},
  \bibinfo{author}{\bibfnamefont{P.}~\bibnamefont{Healey}},
  \bibnamefont{et~al.}, \bibinfo{journal}{IEEE T. Electron. Dev.}
  \textbf{\bibinfo{volume}{55}}, \bibinfo{pages}{2078} (\bibinfo{year}{2008}).


\bibitem[{\citenamefont{Gusynin and Sharapov}(2005)}]{Gusynin2005a}
\bibinfo{author}{\bibfnamefont{V.~P.} \bibnamefont{Gusynin}} \bibnamefont{and}
  \bibinfo{author}{\bibfnamefont{S.~G.} \bibnamefont{Sharapov}},
  \bibinfo{journal}{Phys. Rev. B} \textbf{\bibinfo{volume}{71}},
  \bibinfo{pages}{125124} (\bibinfo{year}{2005}).

\bibitem[{\citenamefont{Beenakker and Vanhouten}(1991)}]{Beenakker1991}
\bibinfo{author}{\bibfnamefont{C.~W.~J.} \bibnamefont{Beenakker}}
  \bibnamefont{and}
  \bibinfo{author}{\bibfnamefont{H.}~\bibnamefont{Vanhouten}},
  \bibinfo{journal}{Solid State Physics} \textbf{\bibinfo{volume}{44}},
  \bibinfo{pages}{1} (\bibinfo{year}{1991}).

\bibitem[{\citenamefont{de~Heer et~al.}(2006)\citenamefont{de~Heer, Berger, and
  First}}]{deHeer2006}
\bibinfo{author}{\bibfnamefont{W.~A.} \bibnamefont{de~Heer}},
  \bibinfo{author}{\bibfnamefont{C.}~\bibnamefont{Berger}}, \bibnamefont{and}
  \bibinfo{author}{\bibfnamefont{P.~N.} \bibnamefont{First}},
  \emph{\bibinfo{title}{Patterned thin films graphite devices and methods for
  making the same}} (\bibinfo{year}{2006}), \bibinfo{note}{{US patent 7015142
  (Provisional filed June 2003, Issued March 21, 2006).}}

\end{thebibliography}

\end{document}